\documentclass[a4paper,11pt]{article}
\usepackage{amssymb,amsmath}
\usepackage{upgreek}
\usepackage{makecell}
\pdfoutput=1 % if you are submitting a pdflatex (i.e. if you have
\usepackage{jinstpub} % for details on the use of the package, please
\title{\boldmath Machine learning technique to improve anti-neutrino detection efficiency for the ISMRAN experiment}

\author[a,b,1]{D.~Mulmule,\note{Corresponding author.}}
\author[a]{P.~K.~Netrakanti,}
\author[a,b]{L.~M.~Pant}%\note{Also at Some University.}}
\author[a,b]{and B.~K.~Nayak}

\affiliation[a]{Nuclear Physics Division, Bhabha Atomic Research Centre,\\ Trombay, Mumbai, India - 400085}
\affiliation[b]{Homi Bhabha National Institute, \\Anushakti Nagar, Mumbai, India - 400094}

\emailAdd{dhruvm@barc.gov.in}

\abstract{The Indian Scintillator Matrix for Reactor Anti-Neutrino detection - ISMRAN experiment aims to detect electron anti-neutrinos ($\overline\nuup_{e}$) emitted from a reactor via inverse beta decay reaction (IBD). The setup, consisting of 1 ton segmented Gadolinium foil wrapped plastic scintillator array, is planned for remote reactor monitoring and sterile neutrino search. The detection of prompt positron and delayed neutron from IBD will provide the signature of $\overline\nuup_{e}$ event in ISMRAN. The number of segments with energy deposit ($\mathrm{N_{bars}}$) and sum total of these deposited energies are used as discriminants for identifying prompt positron event and delayed neutron capture event. However, a simple cut based selection of above variables leads to a low $\overline\nuup_{e}$ signal detection efficiency due to overlapping region of $\mathrm{N_{bars}}$ and sum energy for the prompt and delayed events. Multivariate analysis (MVA) tools, employing variables suitably tuned for discrimination, can be useful in such scenarios. In this work we report the results from an application of artificial neural network - the multilayer perceptron (MLP), particularly the Bayesian extension - MLPBNN, to the simulated signal and background events in ISMRAN. The results from application of MLP to classify prompt positron events from delayed neutron capture events on Hydrogen, Gadolinium nuclei and also from the typical reactor $\gamma$-ray and fast neutron backgrounds is reported. An enhanced efficiency of $\sim$91$\%$ with a background rejection of $\sim$73$\%$ for prompt selection and an efficiency of $\sim$89$\%$ with a background rejection of $\sim$71$\%$ for the delayed capture event, is achieved using the MLPBNN classifier for the ISMRAN experiment.}

\begin{document}
\maketitle
\flushbottom

\section{Introduction}
\label{sec:intro}
In recent years, the measurement of reactor based anti-neutrinos ($\overline\nuup_{e}$) has provided key aspects in understanding the nature of neutrino interactions and their oscillations. Results from experiments such as Double Chooz, Daya Bay and RENO collaborations have reported measurements of $\theta_{13}$ mixing parameter~\cite{DChooz,DBMIX,RENO1} and possibility of searches for light sterile neutrinos~\cite{DBSTR}. Also an excess of events, in energy region $\sim$5 MeV, observed in the prompt event spectrum of the reactor $\overline\nuup_{e}$ induced inverse beta decays (IBD), has opened up new avenues for further studies in reactor based $\overline\nuup_{e}$~\cite{RENO2,DAYABAY,NEOS}. Various experiments, using moderate scale (few tonnes) detector, are being proposed or taking data at very short baselines to further understand the properties associated with the reactor anti-neutrinos~\cite{PROSPRL,STERPRL,SoLid,PANDA}. The Indian Scintillator Matrix for Reactor Anti-Neutrinos (ISMRAN) experiment is one such detector consisting of plastic scintillator (PS) bars in an array forming an active detection volume of 1.0 ton by weight~\cite{ISMREF}. ISMRAN is proposed to measure reactor $\overline\nuup_{e}$s for non-intrusive monitoring of reactor along with the possibility of very short baseline oscillation searches of $\overline\nuup_{e}$ to sterile states.
\begin{figure}%[h]
\begin{center}
\includegraphics[scale=0.40]{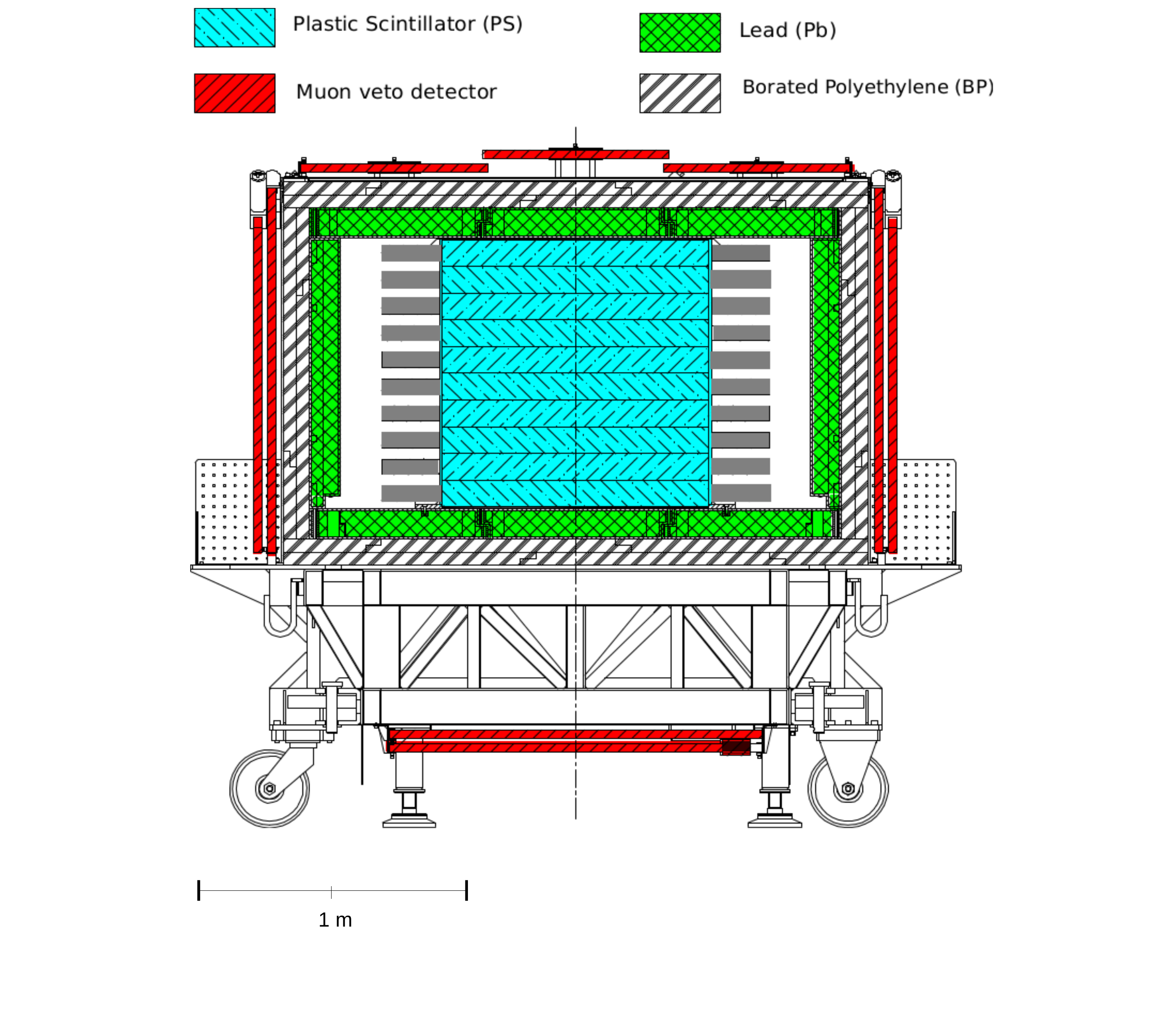}
\caption{Schematic diagram of ISMRAN detector, 100 PS bars, inside a shielding of 10 cm of lead and 10 cm of borated polyethylene on a mobile trolley. Outside the shielding structure are the muon veto scintillator detectors.}
\label{ISMRAN100}
\end{center}
\end{figure}

ISMRAN is an above ground experiment comprising of 100 PS bars arranged in segmented geometry, each of dimension $\mathrm{100 cm \times 10 cm \times 10 cm}$. Each PS bar is wrapped with Gadolinium oxide ($\mathrm{Gd_{2}O_{3}}$) coated aluminized mylar foils (areal density of $\mathrm{Gd_{2}O_{3}}$ : $\mathrm{4.8 mg/cm^{2}}$) and directly coupled to a 3'' PMT at both ends. A 10 cm thick lead (Pb) and 10 cm thick borated polyethylene (BP) shielding will enclose the full setup along with the use of muon veto scintillators on all sides outside the shielding structure. The location of ISMRAN is at a distance of $\sim$13m from a 100 $\mathrm{MW_{th}}$ Dhruva reactor~\cite{DHRUVA} core on a trolley based structure for allowing movement of the detector to various distances from the reactor core. A schematic of complete ISMRAN setup is shown in Fig.~\ref{ISMRAN100}. A high sampling rate (500 MS/s) digitizer based DAQ has been chosen for recording the output signals from the 200 PMT channels and the acquisition will be performed with minimum thresholds for offline event reconstruction. The complete detector system will be housed inside reactor hall and would face a harsh environment of reactor related backgrounds.

The rate of interaction of reactor $\overline\nuup_{e}$ per second $\mathrm{(N_{\overline\nuup_{e}})}$ inside the 1$\mathrm{m^{3}}$ scintillator volume of the ISMRAN detector can be estimated using the below formula:
\begin{equation}
\mathrm{N_{\overline\nuup_{e}} = \frac{N_{p} \cdot P_{th} \cdot \etaup \cdot \overline\sigmaup_{IBD} } {4 \piup D^{2} \cdot \overline E_{f} \cdot 1.6 \cdot 10^{-19}}},
\end{equation}
where the inputs are, $\mathrm{N_{p}}$ : number of quasi-free protons in the scintillator volume, $\mathrm{P_{th}}$ : thermal power of the reactor in MW, $\etaup$ : detection efficiency, $\mathrm{D}$ : distance (in cm) between the detector and center of the core (assuming a compact core), $\mathrm{\overline E_{f}}$ : average energy released per fission in MeV and $\mathrm{\overline\sigmaup_{IBD}=\int\sigmaup(E_{\overline\nuup_{e}})f(E_{\overline\nuup_{e}})d N_{\overline\nuup_{e}}(E_{\overline\nuup_{e}})}$ : the cross section (in $\mathrm{cm^{2}}$) of IBD averaged over the $\overline\nuup_{e}$ spectrum ($\mathrm{E_{\overline\nuup_{e}}}$ is the $\overline\nuup_{e}$ energy and $\mathrm{f(E_{\overline\nuup_{e}})}$ is the $\overline\nuup_{e}$  spectrum per fission). Using this formula the signal rate for ISMRAN at 25$\%$ efficiency comes out to $\sim$100 $\overline\nuup_{e}$ events/day~\cite{ISMREF}.

Feasibility studies using simulations have been performed to evaluate the possibilities for $\overline\nuup_{e}$ detection in ISMRAN setup for the choice of the detector configuration~\cite{varchaswi}. The sensitivity of the chosen setup is extensively studied for the sterile neutrino oscillation search~\cite{Dipak}. It has been calculated that for ISMRAN, a sterile neutrino sensitivity at 90$\%$ C.L. exclusion limits can be achieved with a moderate detection efficiency of 25$\%$ and energy resolution ($\sigma/E$) $\sim$ 20$\%$/$\sqrt(E)$. Improvements in detection efficiency and energy resolution, would help in increasing the sensitivity by almost $\sim$23$\%$. For reactor monitoring, again an efficiency of $\sim$25$\%$ would allow to distinguish $\overline\nuup_{e}$ signal from background at 3$\sigma$ level in $\sim$15 days of ISMRAN reactor ON data, where the background rate of $\sim$1 kHz is expected in full ISMRAN, based on measurements using a prototype shielded setup~\cite{ISMREF}.

In this paper, we present the results from simulations to show the improvements in detection efficiencies of prompt and delayed events by using machine learning technique. A Multilayer Perceptron (MLP) classifier is implemented and a preliminary training of the classifier using information from PS bar hits ($\mathrm{N_{bars}}$) and their sum energy is performed. The prompt and delayed events are further classified from simulated background events consisting of $\gamma$-rays and fast neutron from the surroundings in reactor hall. We also add a new variable $\mathrm{D_{k}}$, apart from the usual sum energy and $\mathrm{N_{bars}}$, which improves the efficiencies of prompt and delayed events in terms of MLP classification. Furthermore, the backgrounds due to cosmic muons and their induced neutrons, long-lived radioactive nuclei, especially $\mathrm{{}^{9}Li/{}^{8}He}$, can mimic a signature of either prompt, delayed or both of IBD candidate events in $\overline\nuup_{e}$ detection experiments~\cite{DCBKG}. Training and development of MLP classifier for such background classification from the true IBD $\overline\nuup_{e}$ events can be explored in future work.
\section{Anti-neutrino detection principle in ISMRAN}
Detection of $\overline\nuup_{e}$s from nuclear reactor is primarily done using the IBD reaction which has a $\overline\nuup_{e}$ energy threshold of 1.806 MeV. In this reaction an $\overline\nuup_{e}$ interacts with a proton in the detector volume (usually a scintillator) and produces positron and neutron (eq.~\ref{eq:ibd}). Due to the reaction kinematics, the positron carries majority of $\overline\nuup_{e}$ energy. The positron deposits energy in the scintillator bars, via ionization, followed by production of two $\gamma$-rays of 0.511 MeV each, from its annihilation with an electron. These $\gamma$-rays then deposit their energy in scintillator bars through Compton scattering. The energy deposited by positron and the resulting annihilation $\gamma$-rays forms the prompt event in ISMRAN detector. While, the neutron having few keVs of energy undergoes thermalization in the detector volume and gets captured on either Gd or H nuclei. The capture of neutron on H will produce a mono-energetic $\gamma$-ray of 2.2 MeV as shown in equation~\ref{eq:HC}. On the other hand, the neutron capture on Gd leads to an emission of cascades of $\gamma$-rays, as shown in equations~\ref{eq:Gd1} and ~\ref{eq:Gd2}, and forms the signature of delayed event. 
\begin{equation}\label{eq:ibd}
  \mathrm{ \overline \nuup_{e} + p \rightarrow e^{+} + n}.
\end{equation}
The capture cross-sections ($\mathrm{\sigmaup_{n-capture}}$) of thermal neutrons on $\mathrm{{}^{155}Gd}$ (eq.~\ref{eq:Gd1}) and $\mathrm{{}^{157}Gd}$ (eq.~\ref{eq:Gd2}) are $\mathrm{61000~b}$ and $\mathrm{254000~b}$, respectively. These cross-sections are about $\mathrm{10^{6}}$ times higher than that for H nuclei. Therefore, after thermalization, the neutron produced from IBD interaction has higher probability to get captured on the wrapped Gd foil outside plastic scintillator bar, than in the hydrogenous bulk of the scintillator in ISMRAN geometry. In recent years, a great amount of experimental and theoretical work is done in understanding and improving the modeling of the de-excitation of Gd nucleus and emission of $\gamma$-ray cascades particularly in context of IBD events~\cite{Yano,Hagiwara,STEREO}.
\begin{equation}\label{eq:HC}
  \hspace{-0.2in}
\mathrm{ n + p \rightarrow d^{*} \rightarrow \gamma,   \quad E_{\gamma} = 2.2~MeV}, \quad \sigmaup_{n-capture} = \mathrm{0.3~b},
\end{equation}
\begin{equation}\label{eq:Gd1}
  \hspace{-0.2in}
\mathrm{ n + {}^{155}Gd \rightarrow {}^{156}Gd^{*} \rightarrow \gamma 's,   \quad \sum E_{\gamma} = 8.5~MeV}, \quad \sigmaup_{n-capture} = \mathrm{61000~b},
\end{equation}
\begin{equation}\label{eq:Gd2}
\hspace{-0.2in}
\mathrm{ n + {}^{157}Gd \rightarrow {}^{158}Gd^{*} \rightarrow \gamma 's,  \quad \sum E_{\gamma} = 7.9~MeV}, \quad \sigmaup_{n-capture} = \mathrm{254000~b}.
\end{equation}
The thermalization and capture of neutron produced from IBD happens after a mean time delay, ranging from a few $\mathrm{\mu}$s up to an order of a few 100 $\mathrm{\mu}$s, from the time of positron event depending on the capture agent position and its concentration in the detector volume. Electron anti-neutrino events in ISMRAN are therefore identified through the detection of these prompt and delayed event pairs separated in time.
\section{Event building in ISMRAN detector}
The ISMRAN detector, positioned at $\sim$13 m from the natural uranium fuel core inside the reactor hall, is a first attempt of its kind for detection of reactor $\overline\nuup_{e}$ in extreme conditions of background. Inside reactor hall, the ambient $\gamma$-ray and neutron background, are expected to be of the order of $\mathrm{\sim10^{3}}$ Hz with 100 PS bars~\cite{ISMREF}. The low IBD cross-section of $\overline\nuup_{e}$ and relatively higher background rates inside reactor hall poses a challenging task of online triggering of the prompt or delayed events. To minimize the losses in triggering of the IBD events, the online data is timestamped and essentially collected in a triggerless mode for all the plastic scintillator bars. The offline analysis consists of event building by grouping of the PS bars ($\mathrm{N_{bars}}$) hit according to the stored timestamps and obtain the sum energy deposition by adding the individual energy deposited in each PS bar. After event building, a classification of an event as either prompt-like or delayed-like needs to be done for further assigning it as a signature of an $\overline\nuup_{e}$ candidate event. At this stage, the information about the mean time delay between an identified prompt-like and delayed-like event may be added as a variable to classify  $\overline\nuup_{e}$ candidate event pairs from other background pairs. Usually, this classification of prompt-like and delayed-like events can be achieved with a selection cut on the sum energy and $\mathrm{N_{bars}}$ variables. However, there is a significant overlap between the individual variable distributions in these event classes which results in reduction of overall $\overline\nuup_{e}$ detection efficiency.
\begin{figure}[h]
\begin{center}
\includegraphics[scale=0.7]{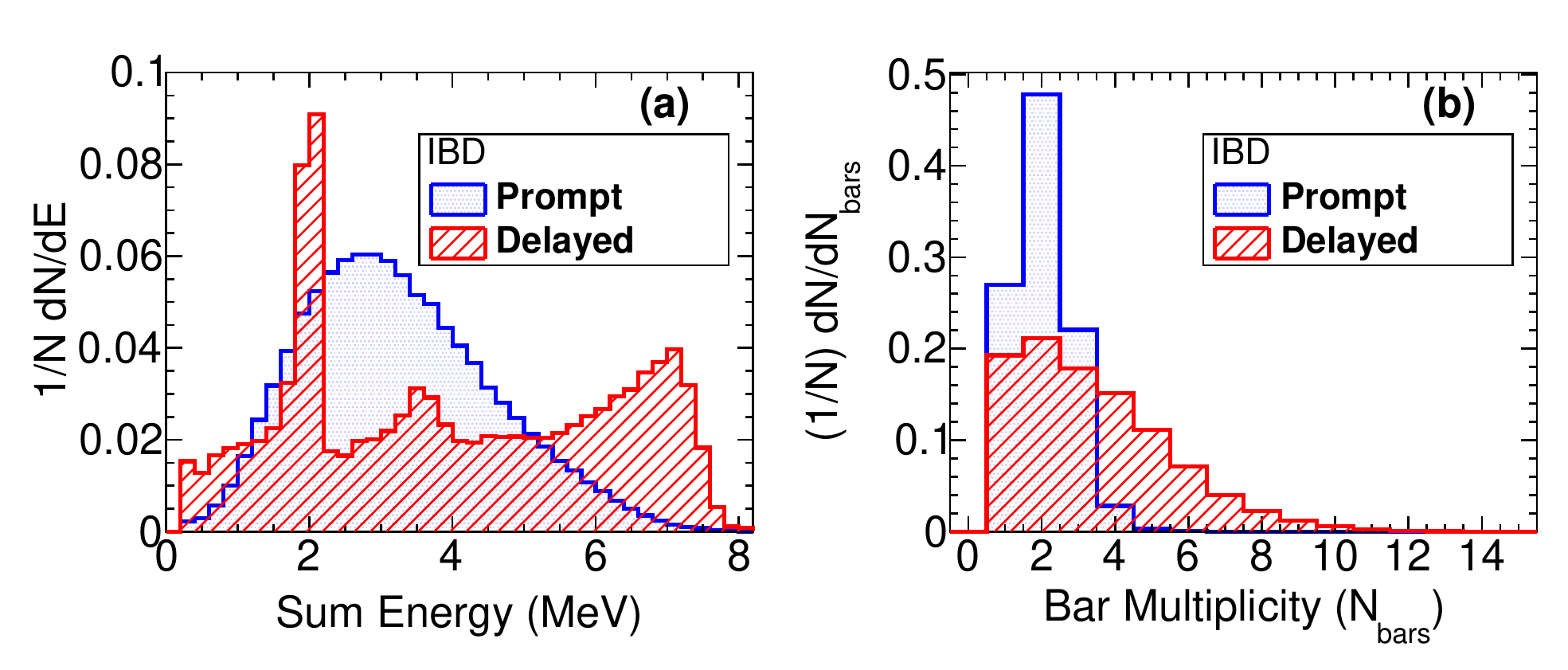}
\caption{Panel (a) Sum Energy and (b) $\mathrm{N_{bars}}$ distributions for prompt and delayed events obtained from Geant4 simulation for IBD events in ISMRAN.}
\label{SumEnBarMult}
\end{center}
\end{figure}
 \section{Simulated IBD events in ISMRAN}
 To understand the IBD event characteristics and the corresponding signal detection efficiencies in ISMRAN, monte-carlo based IBD events are generated using GEANT4~\cite{GEANT4} (version 4.10.4). The physics processes listed in $\mathrm{QGSP\_BIC\_HP}$ are used along with the inclusion of photon evaporation model for the de-excitation of Gadolinium  nucleus and the resulting $\gamma$-ray cascades. The simulations are performed using the parameterization of $\overline\nuup_{e}$ spectrum from ref~\cite{Mueller}, the cross section calculations from ref~\cite{Mention,Vogel} and the fission fractions for different isotopes from ref~\cite{Zhan}.
 The sum energy and $\mathrm{N_{bars}}$ variables are separately recorded for the prompt and delayed events for each simulated IBD interaction in ISMRAN. Only PS bars with deposited energy above the threshold: $\mathrm{E^{Th}_{bar}}$ = 0.2 MeV are considered due to the minimum operating threshold of the signal in the real experiment. Figure~\ref{SumEnBarMult} (a) shows the sum energy and~\ref{SumEnBarMult}(b) the $\mathrm{N_{bars}}$ distributions with threshold condition for both prompt and delayed events. As previously discussed, the overlap in both the distributions for prompt and delayed events is evident in these distributions. An attempt may be made to keep the prompt and delayed event domains exclusive of each other and achieve event classification using the criteria: $\mathrm{N_{bars}^{prompt}}=$  2 or 3 on the prompt event selection and $\mathrm{N_{bars}^{delayed}}>=$ 4, on delayed event selection. The $\overline\nuup_{e}$ signal detection efficiencies obtained for such a simple cut based selection of the prompt and delayed events based on sum energy and $\mathrm{N_{bars}}$ are shown in table~\ref{Table1}, using the ISMRAN simulated IBD events~\cite{ISMREF}. A significant loss in $\overline\nuup_{e}$ detection efficiency is observed for this cut based classification. Moreover, even with these stringent cuts the classification of events as prompt or delayed is still ambiguous. PANDA experimental setup, using an array of 100 plastic scintillator bars have reported detection efficiency of 11.6$\%$ from simulations~\cite{PANDA}.
 \begin{table}[h]
   \begin{small}
     \begin{center}
       \caption{$\overline\nuup_{e}$ detection efficiency with cuts on prompt and delayed events.}
       \label{Table1}
       \begin{tabular}{|c|c|c|c|}
         
         \hline
         \makecell{Loose cuts} & Efficiency ($\%$) & \makecell{Stringent cuts} & Efficiency ($\%$)\\
         \hline
         \makecell{1.8 $<$ $\mathrm{E^{prompt}}$ (MeV) $<$ 8.0, \\ $\mathrm{N_{bars}^{prompt}}$ = 2 or 3}  & 69 & \makecell{2.2 $<$ $\mathrm{E^{prompt}}$ (MeV) $<$ 8.0, \\ $\mathrm{N_{bars}^{prompt}}$ = 2 or 3}  & 67 \\[10pt]
         \hline
         \makecell{0.8 $<$ $\mathrm{E^{delayed}}$ (MeV) $<$ 8.0, \\ $\mathrm{N_{bars}^{delayed}}>=$ 4 }  & 29 & \makecell{3.0 $<$ $\mathrm{E^{delayed}}$ (MeV) $<$ 8.0, \\ $\mathrm{N_{bars}^{delayed}}>=$ 4 }  & 27 \\[10pt]
         \hline
         \makecell{1.8 $<$ $\mathrm{E^{prompt}}$ (MeV) $<$ 8.0, \\ $\mathrm{N_{bars}^{prompt}}$ = 2 or 3 \\ 0.8 $<$ $\mathrm{E^{delayed}}$ (MeV) $<$ 8.0, \\ $\mathrm{N_{bars}^{delayed}}>=$ 4 } & 20 & \makecell{2.2 $<$ $\mathrm{E^{prompt}}$ (MeV) $<$ 8.0, \\ $\mathrm{N_{bars}^{prompt}}$ = 2 or 3 \\3.0 $<$ $\mathrm{E^{delayed}}$ (MeV) $<$ 8.0, \\ $\mathrm{N_{bars}^{delayed}}>=$ 4 }  & 18 \\[25pt]
         \hline
       \end{tabular}
     \end{center}
   \end{small}
 \end{table}

To obtain higher $\overline\nuup_{e}$ detection efficiencies by using the sum energy and $\mathrm{N_{bars}}$ variables, one needs to relax the selection criteria on these variables. This may lead to increase in the false prompt or delayed event rate contribution, from natural backgrounds ($\mathrm{{}^{40}K}$ and $\mathrm{{}^{208}Tl}$), fast neutrons and those from the ambient $\gamma$-ray activity mostly coming from the neutron captures on the material in the vicinity of the detection setup in reactor hall. By performing multivariate analyses (MVA), using the sum energy and $\mathrm{N_{bars}}$ variables, one can enhance the discrimination of the prompt events from the above mentioned backgrounds improving the $\overline\nuup_{e}$ detection efficiency. The event classifiers used in this approach can either be based on multivariate statistics or pattern recognition algorithms. Furthermore, it is possible to utilize additional variables, formed using a weighted combination of base variables, tuned for better discrimination and achieving better signal detection efficiency and purity. Assessment of energy resolution of prompt events from such a classifier will be considered in future studies where a detailed measurement and identification of different sources of the realistic background rates in the reactor hall are available.
\section{Multivariate analysis and artificial neural networks}
Data analysis techniques have evolved extensively in the current phase of high energy physics with powerful techniques available to efficiently extract signal information in background dominated data sample. Techniques using multivariate statistics or machine learning algorithms, such as Maximum Likelihood~\cite{Likelihood1,Likelihood2}, Fischer discriminant~\cite{Fisher}, Boosted Decision Trees~\cite{BDT} or Multilayer Perceptron (MLP)~\cite{MLP1,MLP2}, to separate signal from background events have already been successfully implemented to obtain various interesting physics results. In this work we are going to focus on the use of artificial neural networks (ANN), especially, MLP for the classification of prompt positron signature as signal event against delayed neutron capture and reactor $\gamma$-ray background events.
The multilayer perceptron (MLP), is the traditional form of artificial neutral network~\cite{FrankR}. It is a machine learning algorithm with a structure consisting of an input layer, one or more hidden layers, and an output layer. The MLP algorithm basically approximates a function mapping an $n$-dimensional input $x$ to a $m$-dimensional output $f$ in the real number space. The MLP layers are fully connected i.e. the output of each node is the weighted sum of the outputs of all nodes in the previous layer plus a bias term. These inputs are operated upon by a non-linear sigmoid function like $tanh$ at each node of a hidden layer~\cite{MLP2}. Under certain assumptions, an MLP architecture with a single hidden layer can be shown to approximate any function to arbitrary precision given a sufficient number of hidden nodes~\cite{Hornik,Cybenko}. A supervised learning method~\cite{Rumelhart1} is traditionally used to determine the weights and biases used in an MLP. During this phase, the MLP is presented with data samples where both $x$ and the corresponding output, $f$, referred to as the ground truth, are known, for e.g. from simulations. The measure of the error between the output of the MLP and the ground truth, referred to as the `loss', is computed. An algorithm called the back-propagation algorithm~\cite{Rumelhart2} calculates the gradient of this loss as a function of the weights and biases, which is then minimized by altering the weights and biases using the stochastic gradient descent method~\cite{YLeCun}. Repeated application of the above procedure is carried out till the errors are reduced to an acceptable level. However, in order to reduce the number of iterations to cut down on the computation time an alternative approach called the Broyden-Fletcher-Goldfarb-Shannon (BFGS) method can be utilized while adapting the synapse weights~\cite{BFGSRef}. This method uses the second derivatives of the error function for adjusting the weights in each iteration. 
In this paper, we present the results of prompt signal and delayed background classification for ISMRAN detector by using the `Bayesian' extension of MLP with the above BFGS method incorporated in it, referred to as - `MLPBNN' in the ROOT TMVA~\cite{tmva} package. The MLPBNN approach allows for increasing the complexity (more hidden units and/or more layers) of the architecture while simultaneously employing a regulator to avoid over-training. This is achieved through addition of another term in the network error function that effectively penalizes large weights, consequently controlling the complexity of the model. For purposes of brevity in writing and also acknowledging the fact that MLPBNN is an extension of the more fundamental MLP algorithm, we will use only the term `MLP' for the MLPBNN classifier here onwards. Similar work adopting the convolution visual network from the machine learning methods is used to describe neutrino interactions based on their topology~\cite{Aurisano}.
\section{Application of MLP to ISMRAN event classification}
The MVA based classification, presented in this work, uses a simulated sample of 5M IBD events in ISMRAN detector. PS bars with energy deposit above the threshold : $\mathrm{E^{Th}_{bar} = 0.2}$  MeV and $\mathrm{N_{bars} >}$ 1 are used to reconstruct prompt and delayed events.
The effectiveness of MLP as the classifier of choice for ISMRAN, is first demonstrated against the traditional `likelihood' method. While the method of `cuts' is a series of selections on the variables as attempted in our preliminary approach, the likelihood is a statistical method useful when looking for signal from a limited data sample. Likelihood combines the probability density estimators for each variable and uses this product to classify the sample events. Although the process of both cuts and likelihood methods is transparent, the performance of these classifiers maybe hampered when there is significant correlation among the variables. 

Firstly, we compare the relative performances of likelihood and MLP classifiers for separating prompt positron events from delayed events of neutron capture on either Gd or H nucleus. This classification is foremost and essential to tag reconstructed events in real data as prompt or delayed events before assigning the pairs as $\overline\nuup_{e}$ candidate event. The reconstructed positron prompt events from the $\mathrm{N_{bars}}$ and sum energy deposition are defined as ``Reco prompt'' events. The neutron capture delayed events on Gd comprises of $\sim$75$\%$ of total neutron captures in ISMRAN detector~\cite{ISMREF} and the remaining $\sim$25$\%$ takes place on hydrogen nuclei in the bulk of the scintillator volume. All such neutron capture delayed events which can be misidentified as prompt events are defined as ``False prompt''. The choice of this terminology is made to stress the fact that, the emphasis is on accurately classifying the prompt event, as it is crucial to derive the $\overline\nuup_{e}$ energy distribution. The neutron capture event on the other hand may not be specific to the IBD neutron event. A similar signature as an IBD delayed event is expected when a fast neutron produced from reactor surroundings enters ISMRAN, thermalizes in PS bars and gets captured on Gd or H. A detailed measurement of fast neutron background in reactor hall is required to simulate such fast neutron background event rate in ISMRAN and is not currently available. Hence, for this study, we are considering the simulated IBD neutron capture delayed events as background, since the final cascade $\gamma$-ray signature would be same in both the cases. Since, these fast neutron events will be uncorrelated to the reconstructed prompt event, the mean time delay variable will not be effective to discriminate such background pairs from the real IBD event pairs.
For both methods 100000 events are used for the classifier training. Another set of 100000 events, completely different from the training set, are simultaneously used for testing and evaluation purposes. In the case of MLP, additional inputs such as neuron type, the number of hidden layers, number of neurons in a hidden layer, testing iterations and frequency of the tests are provided to the classifier. 
\begin{figure}[h]
\begin{center}
\includegraphics[scale=0.75]{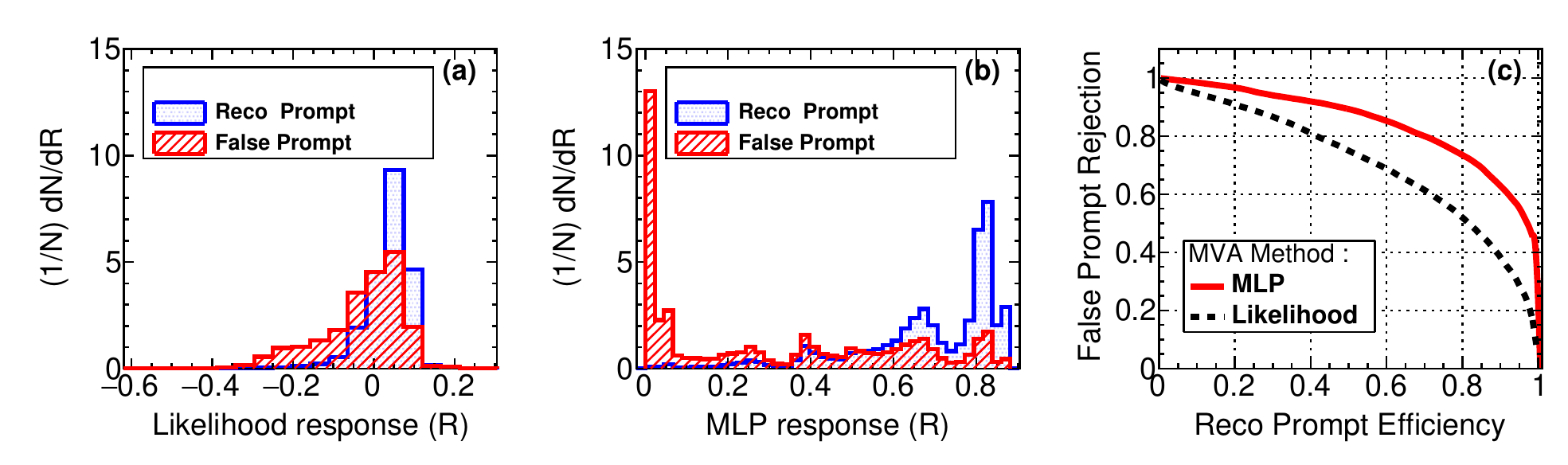}
\caption{Panel (a) and (b) shows the comparison of performance for likelihood and MLP classifiers on simulated IBD reco prompt and false prompt events in ISMRAN detector, respectively. Panel (c) shows the comparison of ROC curve for the MLP and likelihood classifiers.}
\label{Comp_Lkl_MLP}
\end{center}
\end{figure}

The classifier response (R) of likelihood and MLP methods are presented in Fig.~\ref{Comp_Lkl_MLP} (a) and (b), respectively. The separation between the reco prompt and false prompt events for MLP is better than likelihood classifier. Figure~\ref{Comp_Lkl_MLP} (c) shows the reco prompt efficiency vs false prompt rejection curve shows the `receiver operator characteristics' - ROC for the two classifiers. A ROC curve can provide an optimal working point in terms of true positive and false positive selection rates for any predictive model, which could be applied to the data, chosen for classification of events. The ROC curve shows better false prompt background rejection in case of MLP, even for higher reco prompt efficiencies. These results make MLP classifier a better choice for ISMRAN IBD events out of the two compared classifier methods. It must be pointed out that the chosen MLP architecture in our study uses two hidden layers. The first hidden layer uses $N$+5 nodes while the second one uses $N$ nodes, where $N$ corresponds to the number of input variables. One of the input node apart from the input variables is the bias node, which is implicit in the MLP architecture. The two hidden layer configuration is found to have optimal performance for reasonable number of iterations in error minimization leading to less computational time.
\section{Results from MLP classification}
{\bf MLP response for reco prompt and false prompt events for IBD interactions in ISMRAN}\\
Thermal neutron capture on Gd nucleus is followed by emission of $\gamma$-ray cascades from its de-excitation. These cascades of $\gamma$-rays mostly span multiple PS bars and hence the average $\mathrm{N_{bars}}$ in such events would be higher than a mono energetic $\gamma$-ray emission by neutron capture on H nuclei. Also, the sum of these deposited energies in PS bars is expected to be around 8 MeV, for a fully contained event. A selection cut on the MLP classifier, using only sum energy and $\mathrm{N_{bars}}$ variable, which selects $\sim$90$\%$ of reco prompt events from the sample can only provide $\sim$65$\%$ rejection of the false prompt events, as shown in Fig.~\ref{Comp_Lkl_MLP}(c). 
\begin{figure}[h]
\begin{center}
\includegraphics[scale=0.75]{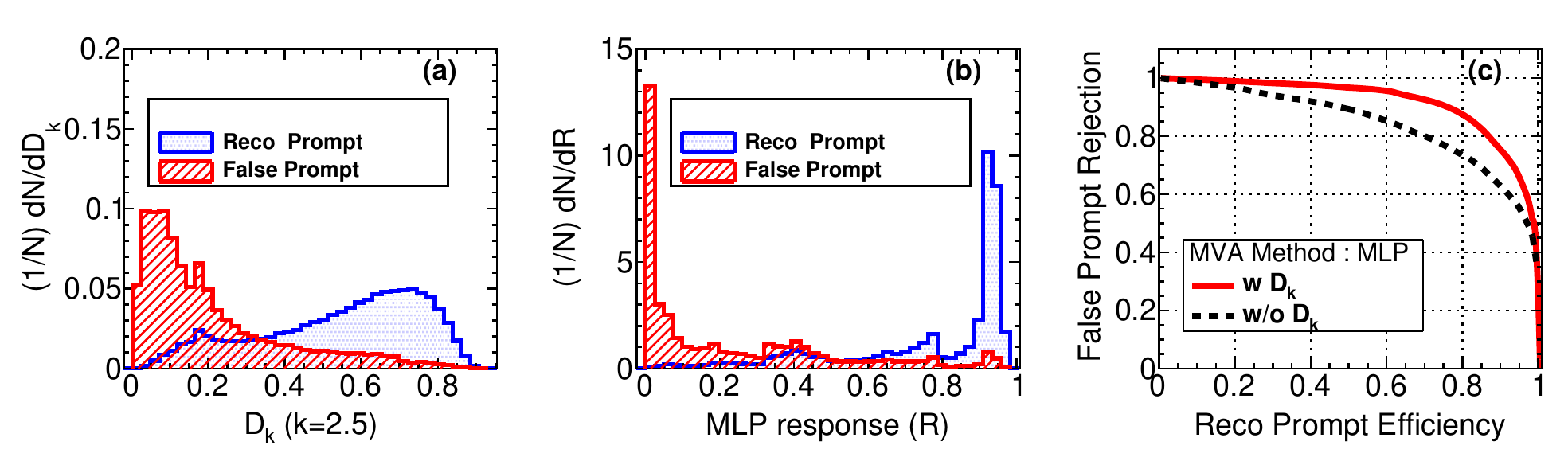}
\caption{Panel (a) shows the reco prompt and false prompt event separation for $\mathrm{D_{k}}$ variable for IBD events in ISMRAN. Panel (b) shows the response after application of MLP classifier including the $\mathrm{D_{k}}$ variable along with $\mathrm{N_{bars}}$ and sum energy variable. Panel (c) shows the improvement in the ROC curve of MLP classifier with inclusion of $\mathrm{D_{k}}$ variable for reco prompt efficiency and false prompt rejection.}
\label{Dk2p5_ROC_Clsfr}
\end{center}
\end{figure}
Further improvements in the existing framework are therefore needed to increase the reco prompt efficiency of the classifier response. In order to achieve this, we introduce a new variable constructed using the weighted individual bar energy deposits formulated as: $\mathrm{D_{k} = (E_{total})^{-k}\times (\sum_{i}(w_{i}\times(E_{i})^{k}))}$, where $\mathrm{E_{total}}$ is the total sum energy, $\mathrm{E_{i}}$ is the individual PS bar energy deposit, $\mathrm{k}$ is a real number, and the weight factor is defined as $\mathrm{w_{i}=E_{i}/E_{total}}$. This additional variable is inspired by discrimination variables used in quark and gluon jet identification in high energy proton-proton collisions~\cite{QGJet}. In case of $\mathrm{D_{k}}$, the choice of exponent $\mathrm{k}$ as 2.5 is observed to provide better discrimination ability compared to other values of $\mathrm{k}$. The formulation of the variable $\mathrm{D_{k}}$ is such that it makes use of the difference in the energy deposition profiles of the reco prompt events and false prompt events and consequently enhances their separation as seen in Fig.~\ref{Dk2p5_ROC_Clsfr}(a). Figure~\ref{Dk2p5_ROC_Clsfr}(b) shows better separation in the MLP classifier response, after inclusion of $\mathrm{D_{k}}$ and Fig.~\ref{Dk2p5_ROC_Clsfr}(c) shows the comparison of ROC curves with and without inclusion of $\mathrm{D_{k}}$ for reco prompt efficiency and false prompt rejection. 
It can be seen that there is a significant improvement in false prompt rejection for a given reco prompt efficiency with the inclusion of $\mathrm{D_{k}}$ in the MLP classifier. All the MLP classifier results presented, here onwards, are with the inclusion of $\mathrm{D_{k}}$ along with sum energy and $\mathrm{N_{bars}}$ variables.\\
{\bf MLP response for reco prompt events and false prompt events from neutron capture on Gd and H separately in ISMRAN}\\
\begin{figure}[h]
\begin{center}
\includegraphics[scale=0.75]{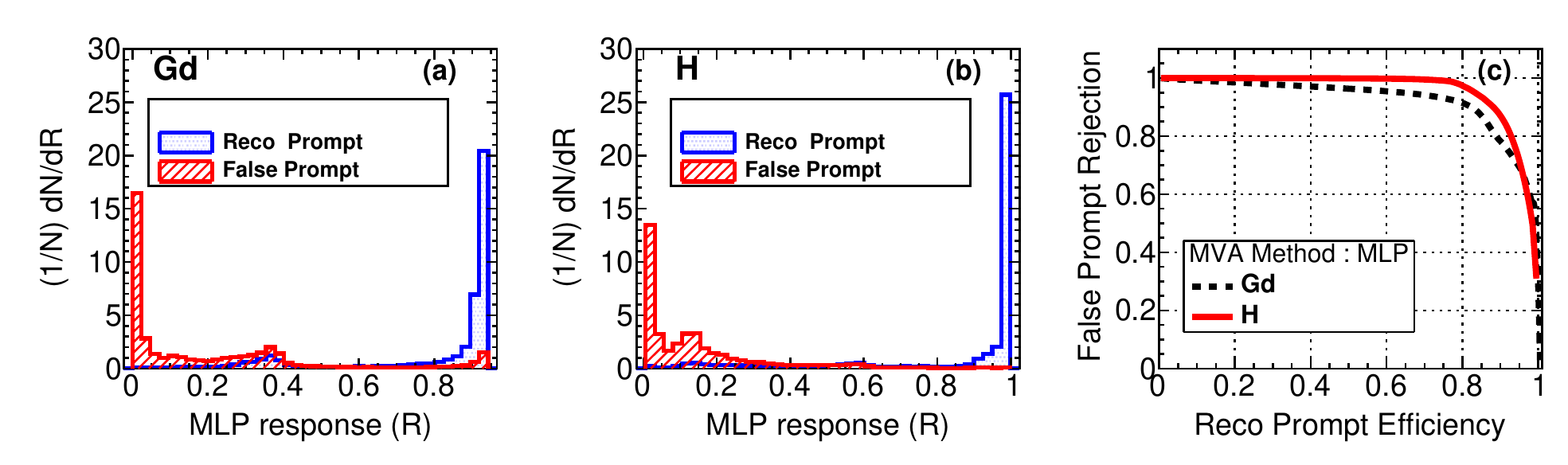}
\caption{Panel (a) and (b) shows the MLP classifier response for reco prompt events from positron and false prompt events from neutron capture on Gd and H, respectively. Panel (c) shows the comparison of ROC curve for the reco prompt efficiency and false prompt rejection from neutron capture events on Gd and H.}
\label{PromptHCap}
\end{center}
\end{figure}
It is important to evaluate the MLP classifier response, separately, to the false prompt events from Gd and H capture delayed events and to obtain the corresponding rejection rates. This is required, because both nuclei undergo de-excitation process leading to different sum energy and $\mathrm{N_{bars}}$ signatures. The motive behind studying classification performance for H capture events and comparing it to Gd captures is to evaluate and ensure that the MLP framework performs equally effectively in both scenarios.This is addressed within the MLP framework by individually studying the false prompt events from neutron capture on Gd and H. Figure~\ref{PromptHCap} (a) and (b) show the MLP response and classification between a reco prompt from positron and false prompt event for neutron capture on Gd and H, respectively. The separation in the classifier response is good in both the events from Gd and H capture events from the reco prompt events. Figure~\ref{PromptHCap} (c) shows the false prompt event rejection due to the Gd and H capture events as a function of reco prompt efficiencies. It can be seen that the rejection of false prompt events using MLP is quite effective in case of both Gd and H capture events keeping reasonable reco prompt efficiency.\\
{\bf MLP response for reco prompt events and reactor $\gamma$-ray background events in ISMRAN}\\
\begin{figure}[b]
\begin{center}
\includegraphics[scale=0.6]{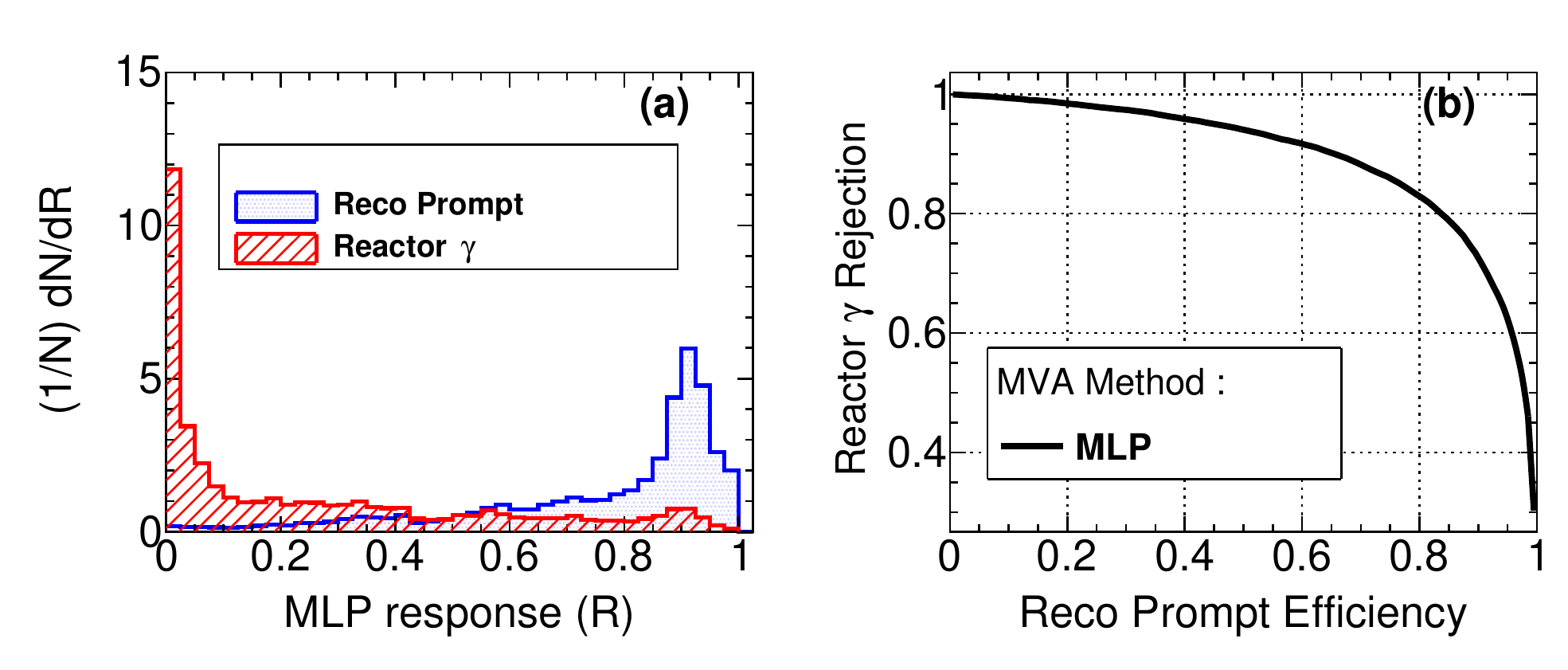}
\caption{Panel (a) MLP classifier response for reco prompt events from positron and reactor $\gamma$-ray backgrounds. Panel (b) ROC curve for the reco prompt efficiency and reactor $\gamma$-ray background rejection.}
\label{PROSPECT}
\end{center}
\end{figure}
The experimental setup of ISMRAN detector inside reactor hall poses a hostile environment for anti-neutrino detection due to the ambient reactor $\gamma$-ray backgrounds. These $\gamma$-rays are emanating mostly from the neutron capture on the surrounding materials present in the reactor hall, namely the stainless steel structures and beam dumps used in reactor operations or for various neutron scattering experiments. We have taken a reference $\gamma$-ray spectrum for such events from the PROSPECT experimental site selection studies~\cite{PROSPECT}, particularly at the  NBSR site where the $\gamma$-ray activity is quite intense and varied in energy. This allows for the test of the MLP classifier response in a more realistic reactor $\gamma$-ray background. The above reactor background $\gamma$-ray distribution is used as an input in our GEANT4 simulations and events are recorded for a shielded ISMRAN detector geometry. These events are considered as the background events, and the MLP classification is trained for discriminating the reco prompt events. The reco prompt event and reactor $\gamma$-ray background event separation are shown in the Fig.~\ref{PROSPECT}(a) along with the ROC curve in Fig.~\ref{PROSPECT}(b). A reco prompt efficiency of $\sim$90$\%$ is achieved with $\sim$70$\%$ of reactor related $\gamma$-ray background event rejection.\\
\begin{figure}[h]
\begin{center}
\includegraphics[scale=0.6]{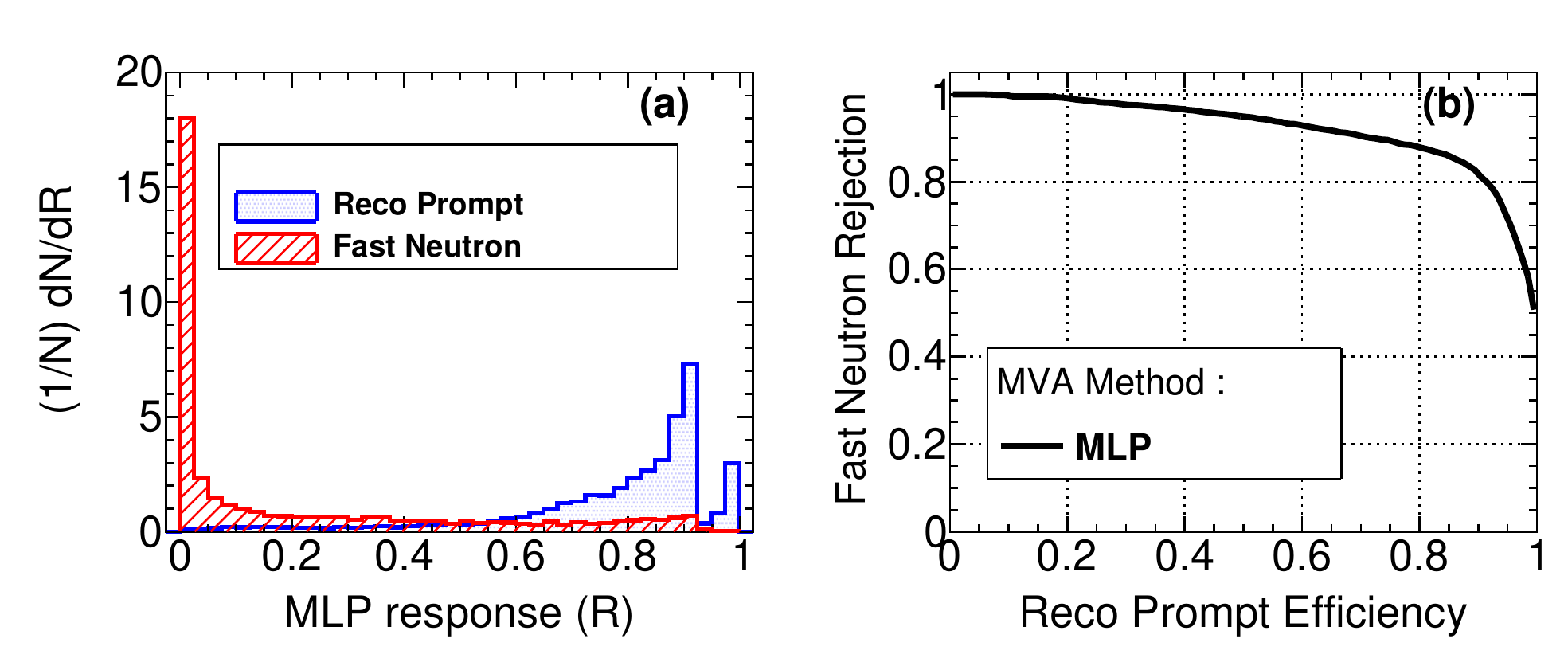}
\caption{Panel (a) MLP classifier response for reco prompt events from positron and proton recoil events from fast neutrons. Panel (b) ROC curve for the reco prompt efficiency and fast neutron rejection.}
\label{FastNeutron}
\end{center}
\end{figure}
{\bf MLP response for reco prompt events and fast neutron background events in ISMRAN}\\
A fast neutron, produced either from cosmogenic source, muon spallation from shielding material or from reactor background, can enter ISMRAN detector and produce an event where a large fraction of neutron energy is transferred to a proton inside the PS bar. Such proton recoil events usually tend to have signatures very close to the reco prompt event from positron. To simulate such proton recoil events a uniform distribution of fast neutrons from 2 MeV to 20 MeV were generated in GEANT4 inside the ISMRAN setup. The sum energy deposition, $\mathrm{N_{bars}}$ and $\mathrm{D_{k}}$ variables from such proton recoil events are then classified from the reco prompt events using MLP. Figure~\ref{FastNeutron} (a) shows the MLP response of reco prompt events and proton recoil events from fast neutrons. The separation in terms of classifier response for these events is reasonably good. Figure~\ref{FastNeutron} (b) shows the ROC curve for the reco prompt efficiency and rejection of proton recoil events from fast neutron. A reco prompt efficiency of $\sim$80$\%$ is achieved with a rejection of proton recoil events close to 84$\%$.\\
{\bf Performance evaluation of MLP for reco prompt and delayed events in ISMRAN}\\
\begin{figure}[t]
\begin{center}
\includegraphics[scale=0.4]{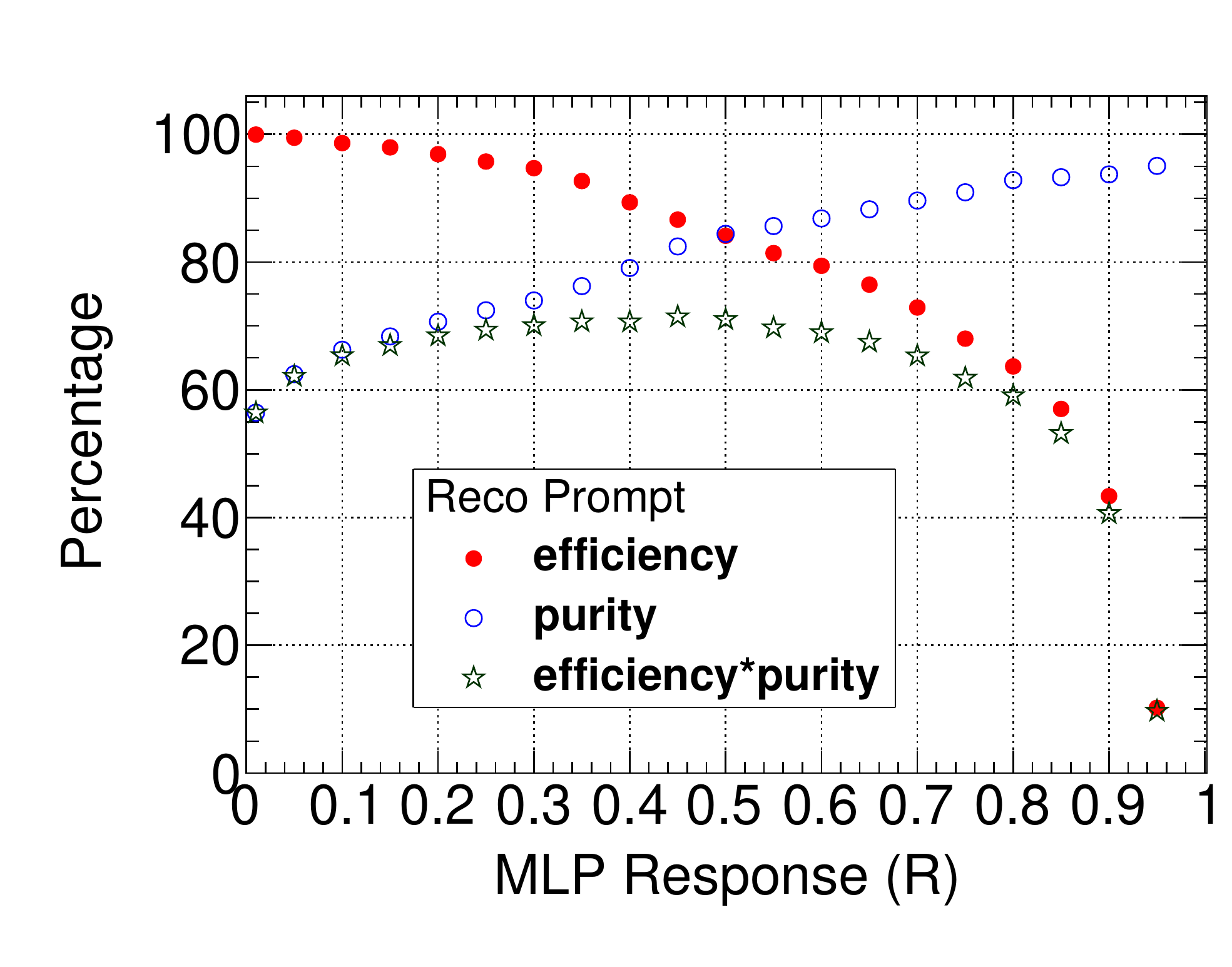}
\caption{Reco prompt efficiency, purity and product of efficiency and purity for MLP as a function of different selection cut values on the MLP output.}
\label{newFinal}
\end{center}
\end{figure}
To test the performance of MLP classifier, we prepare a sample of 100000 events of reco prompt events with a mixed sample of false prompt events from Gd and H neutron capture delayed events. The inclusion of H neutron capture events as potential delayed event is one of the advantages of using the MLP method over the cut based selection which rejected such events and led to reduced $\overline\nuup_{e}$ signal detection efficiencies. The MLP performance parameters such as reco prompt efficiency, purity and their product are calculated as a function of different selection of cut values on the MLP response. Figure~\ref{newFinal} shows the relative behavior of these three parameters for different selection cuts on the MLP classifier in percentages.  In the region of MLP response values from 0.2 to 0.6 the product of signal efficiency and purity is close to 70$\%$. The performance of the MLP classifier is obtained in terms of figures of merit (FOM), $\mathrm{s / \sqrt{(s + b)}}$ and $\mathrm{s / \sqrt{b}}$, where $\mathrm{s}$ and $\mathrm{b}$ define the reco prompt signal and false prompt background events, respectively, in the sample. The quantity $\mathrm{s / \sqrt{(s + b)}}$ can be used to maximize the efficiency of the reco prompt signal events with the classifier response and $\mathrm{s / \sqrt{b}}$ is used for obtaining the maximum purity of the reco prompt signal in presence of false prompt background events.
\begin{figure}[b]
\begin{center}
\includegraphics[scale=0.4]{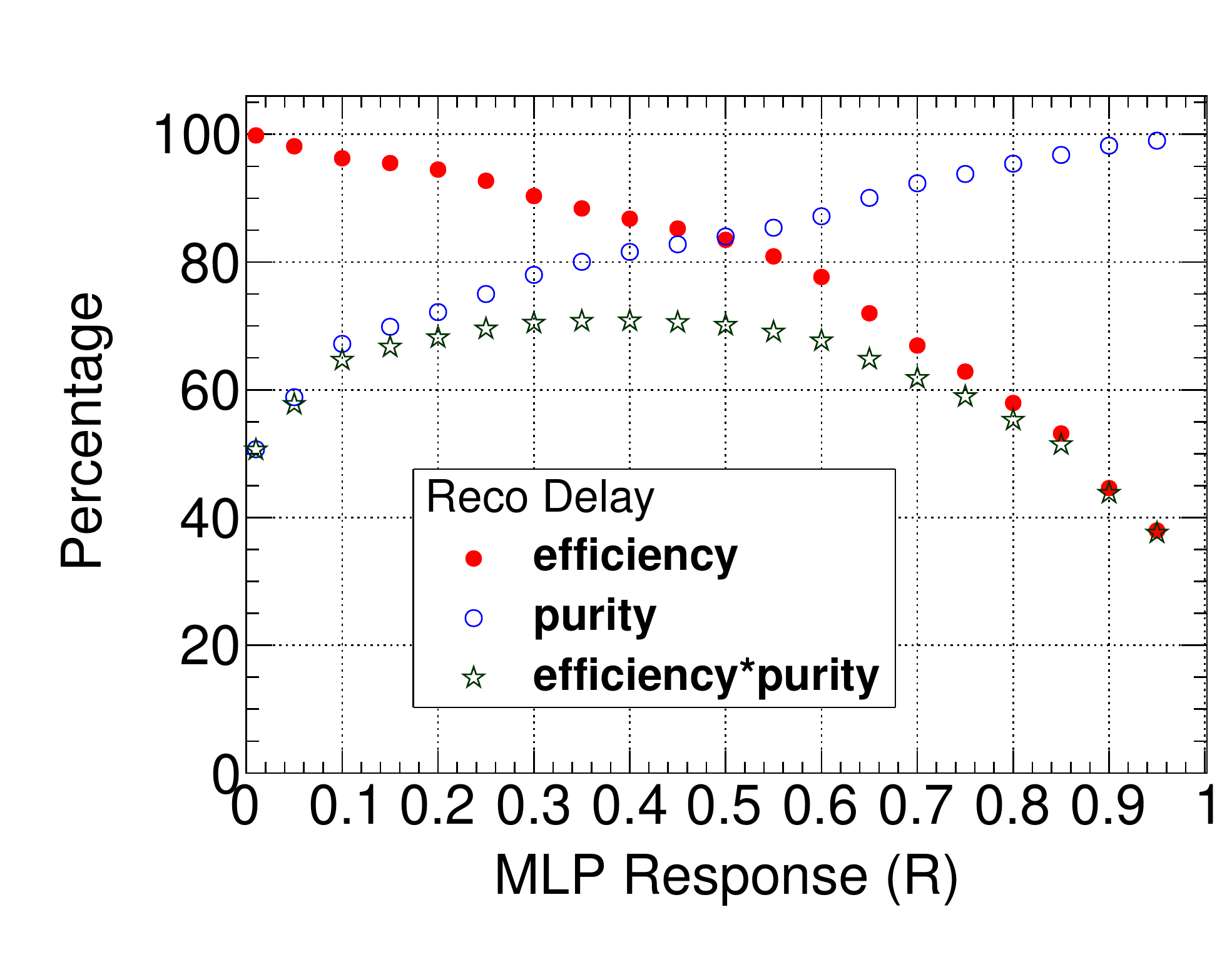}
\caption{Reco delay efficiency, purity and product of efficiency and purity for MLP as a function of different selection cut values on the MLP output.}
\label{newFinalDelay}
\end{center}
\end{figure}
Similar analysis using MLP, is also performed for an IBD neutron capture delayed event under classification. Here, the false delayed events are those reco prompt events which may have similar signature of IBD neutron capture delayed events in terms of sum energy deposition,  $\mathrm{N_{bars}}$ and $\mathrm{D_{k}}$ variables. Figure~\ref{newFinalDelay} shows the efficiency, purity and product of efficiency and purity for the reco delayed events. The efficiency, purity and false rejection for reco prompt and delayed events obtained for specific cut values on MLP response are tabulated in table~\ref{Table2a} and~\ref{Table2b} respectively. These selection cut values are chosen, so as to maximize the $\mathrm{s / \sqrt{(s + b)}}$ and $\mathrm{s / \sqrt{b}}$ of both event classes and provide an evaluation of the MLP performance.
%-------------------Table 2a starts-------------------------
\begin{table}
\begin{small}
  \begin{center}
  \caption{Efficiency, purity and false rejection performance for reco prompt events.}
  \label{Table2a}
\begin{tabular}{|c|c|c|c|c|}
\hline
                           & MLP cut value   & Efficiency ($\%$)        & Purity ($\%$)               & False Rejection ($\%$)\\
\hline
$\mathrm{s /\sqrt{s+b}}$   & 0.37            & 91.5                     & 77.3                        & 73.1\\
\hline
$\mathrm{s /\sqrt{b}}$     & 0.88            & 56.4                     & 93.5                        & 96.1\\
\hline
\end{tabular}
\end{center}
\end{small}
\end{table}

\begin{figure}[h]
\begin{center}
\includegraphics[scale=0.6]{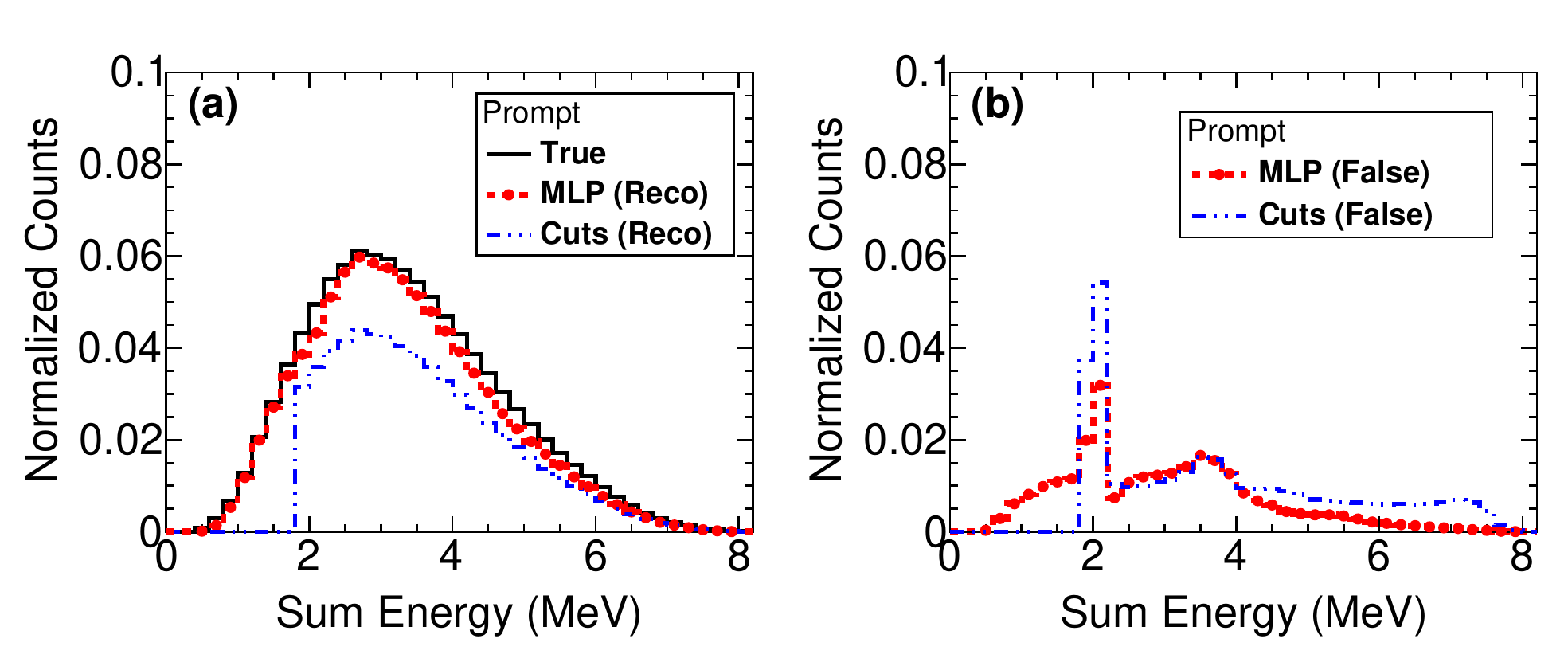}
\caption{Panel(a) : Prompt sum energy distribution for true input, reco prompt events classified with MLP and from a cut based analysis. Panel(b) shows the false prompt events which are misidentified as reco prompt events using MLP classifier and cut based analysis.}
\label{CrossCheckMLP}
\end{center}
\end{figure}

%--------------------Table 2a ends and 2b starts-------------------------
\begin{table}
\begin{small}
  \begin{center}
  \caption{Efficiency, purity and false rejection performance for reco delayed events.}
  \label{Table2b}
\begin{tabular}{|c|c|c|c|c|}
\hline
                           & MLP cut value   & Efficiency ($\%$)        & Purity ($\%$)               & False Rejection ($\%$)\\
\hline
$\mathrm{s /\sqrt{s+b}}$   & 0.37            & 88.7                     & 80.5                        & 71.3\\
\hline
$\mathrm{s /\sqrt{b}}$     & 0.70            & 66.9                     & 92.3                        & 94.4\\
\hline
\end{tabular}
\end{center}
\end{small}
\end{table}
%--------------------Table 2b ends-------------------------
The efficiency value, shown in table~\ref{Table2a}, of 91.5$\%$, as obtained for maximally efficient classification of reco prompt events, is a significant gain over the earlier cut based efficiency of 69$\%$(see table~\ref{Table1}). The maximum purity of the reco prompt events can reach $\sim$93$\%$ at the MLP selection cut of 0.88 with an efficiency of $\sim$56$\%$ for the reco prompt events. The optimal selection of cut value on the MLP response may be selected in these ranges to achieve a moderate reco prompt signal efficiency with a reasonable reco prompt signal purity. Similarly, table~\ref{Table2b} presents the maximum possible reco delayed event efficiency of 88.7$\%$ obtained at a cut of 0.37, with a purity of 80.5$\%$ and false rejection of 71.3$\%$. For the selection of $\mathrm{s /\sqrt{b}}$,  a optimal cut of 0.70, selects the reco delayed events with 92.3$\%$ purity with a false delayed rejection of 94.4$\%$. The achieved efficiency, purity and false rejection obtained from MLP classifier for reco prompt and reco delayed events are  the preliminary estimations and provides a better performance than the results obtained from a simple cut based analysis.
The final $\overline\nuup_{e}$ energy spectra is closely related to the measured reco prompt energy distribution. To evaluate the spectral shape, the performance of the MLP classifier on selecting the reco prompt signal events is tested using 100000 independent IBD events.  All the events which satisfies the classifier cut selection above 0.37 are chosen as reco prompt signal events. Figure~\ref{CrossCheckMLP} (a) shows the prompt sum energy distribution of input events (shown in black), after application of the MLP classifier response on the reco prompt events  (shown in red dashed) and that from a simple cut based analysis choosing `loose' cuts on reco prompt events, as in table~\ref{Table1}(shown in dashed dot blue). It can be seen that using MLP a significant improvement in the efficiency and spectral shape of prompt energy distribution is obtained as compared to a simple cut based analysis results. A very conservative systematic uncertainty is evaluated due to the MLP response variations. The source of these systematic uncertainties arise primarily from the choice of input model used for training and testing of the MLP classifier, incorporation of all the associated backgrounds in the reactor hall and due to the variation in the efficiencies of the detector over the entire data taking duration. We varied the MLP response cut value by 5$\%$ and estimated the uncertainty of 2.4 $\%$ in the reco prompt energy distribution.
Figure~\ref{CrossCheckMLP} (b) shows the false prompt energy distribution for MLP and cut based analysis which are filtered as reco prompt signal events. It can be seen that the MLP based classifier are accepting less number of false prompt events which can be marked as reco prompt signal events as compared to that from a cut based analysis. Also more number of background events, at around 2MeV from neutron capture on H, are misidentified as reco prompt signal events in case of cut based analysis as compared with MLP classifier. Overall the purity of the reco prompt signal events is higher in case of MLP classifier as compared to a simple cut based analysis.

\section{Conclusions and Outlook}
Machine learning technique using multilayer perceptron algorithm is applied to discriminate reco prompt events arising from IBD interactions from the false prompt events from different sources of background in the ISMRAN detector. Using simulations, it has been shown that this technique provides an excellent separation of reco prompt events from false prompt events which are from the delayed neutron capture on Gd, H, fast neutron and $\gamma$-rays background from reactor. The performance of MLP classifier is better as compared to the statistical methods of cut based or likelihood based classification. An addition of new variable, $\mathrm{D_{k}}$, obtained from weighted energy deposits in bars further improves the response of MLP classifier. Prompt signal efficiencies close to $\sim$91$\%$ has been obtained while rejecting $\sim$73$\%$ of the false prompt events in ISMRAN detector. In future, MLP classifier may be used to discriminate other source of backgrounds which include cosmogenic muon spallation and neutron induced backgrounds for obtaining better anti-neutrino detection efficiency in ISMRAN.

\end{document}